\documentclass[10pt, conference, letterpaper]{IEEEtran}
\usepackage{graphicx}
\usepackage{epstopdf}
\usepackage{cite}
\usepackage{amsfonts}
\usepackage{amssymb}
\usepackage{amsmath}
\usepackage{mathrsfs}
\usepackage{dsfont}
\usepackage{bm}
\usepackage{amsmath}
\usepackage{mdwmath}
\usepackage{mdwtab}
\usepackage{url}
\usepackage[ruled,vlined,linesnumbered]{algorithm2e} 
\usepackage{subfigure}
\allowdisplaybreaks

\newtheorem{remark}{Remark}
\newtheorem{theorem}{Theorem}
\newtheorem{lemma}{Lemma}

\newtheorem{defi}{Definition}
\newtheorem{coro}{Corollary}

\newenvironment{iarray}{\begin{IEEEeqnarray}{rCl}}{\end{IEEEeqnarray}\ignorespacesafterend}

\usepackage{makecell,multirow,diagbox}
\newcommand{\tabincell}[2]{\begin{tabular}{@{}#1@{}}#2\en
		d{tabular}}

\begin{document}

\title{Round-Robin is Provably Near-Optimal for Minimizing Age with HARQ over Heterogeneous Unreliable Multiaccess Channels
}
\author{\IEEEauthorblockN{Zhiyuan Jiang}
	\IEEEauthorblockA{Shanghai Institute for Advanced Communication and Data Science, Shanghai University, Shanghai 200444, China.\\
	Email: jiangzhiyuan@shu.edu.cn
}
}
\maketitle

\begin{abstract}
In a heterogeneous unreliable multiaccess network, wherein terminals share a common wireless channel with distinctive error probabilities, existing works have showed that a persistent round-robin (RR-P) scheduling policy (i.e., greedy policy) can be arbitrarily worse than the optimum in terms of Age of Information (AoI) under standard Automatic Repeat reQuest (ARQ), and one must resort to Whittle's index approach for optimal AoI. In this paper, practical Hybrid ARQ (HARQ) schemes which are widely-used in today's wireless networks are considered. We show that RR-P is very close to optimum with asymptotically many terminals in this case, by explicitly deriving tight, closed-form AoI gaps between optimum and achievable AoI by RR-P. In particular, it is rigorously proved that for RR-P, under HARQ models concerning fading channels (resp. finite-blocklength regime), the relative AoI gap compared with the optimum is within a constant of $(\sqrt{e}-1)^2/4\sqrt{e} \cong 6.4\%$ (resp. $6.2\%$ with error exponential decay rate of $0.5$). In addition, RR-P enjoys the distinct advantage of implementation simplicity with channel-unaware and easy-to-decentralize operations, making it favorable in practice.
\end{abstract}

\section{Introduction}
\label{sec_intro}
Age of Information (AoI) \cite{kaul12} has been extensively studied in recent years, which represents the time elapsed since the generation of the newest received status at a destination. Compared with conventional end-to-end communication delay metrics, AoI captures the timeliness of information and is intrinsically related to the effectiveness of critical status information, e.g., in connected autonomous driving and time-sensitive networked control systems. Therefore, there is a growingly strong motivation of optimizing AoI in wireless networks, as the future wireless systems are more and more concerned with machine-type applications.

In wireless systems, one of the most prevalent and representative settings can be modeled as unreliable multiaccess networks, wherein terminals share a common wireless channel (error-prone due to noise and channel fading) to communicate with a master node, e.g., a central controller or base station. Kadota \emph{et. al} \cite{kadota18} considered the independently identically distributed (i.i.d.) channels with terminal-dependent error probabilities, standard Automatic Repeat reQuest (ARQ) and active sources (i.e., sources generate a fresh status whenever scheduled), in which case AoI is in fact identical with the definition of time-since-last-service in \cite{ery13}. It was shown that, intuitively, the optimal policy that minimizes the time-average AoI should serve the terminals both timely and regularly. The timeliness requirement is related to the first-order metric which is known as peak-age (i.e., equivalent with inter-delivery time in \cite{idt15}); and the regularity requirement is related to the second-order moment of peak-age. A stationary randomized policy with optimized access probabilities is proved that can minimize the peak-age \cite{talak_mobihoc}, but only $2$-optimal which means its AoI is within twice the optimum due to its non-regularity. A Persistent Round-Robin (RR-P) policy (i.e., greedy policy in \cite{kadota18} since it selects the terminal with highest current AoI) is as regular as it can be, but falls short for peak-age and is shown to not have a constant multiplicative optimality guarantee. In the literature, it is found that a Whittle's index policy \cite{jiang19_tcom,hsu18,ali20} is practically and asymptotically optimal with many terminals since it schedules a terminal based on a scaled age, i.e., approximately $\sqrt{p_i}{h_i}$ where $p_i$ and $h_i$ are success probability and AoI of terminal-$i$ respectively, which jointly accounts for timeliness and regularity in an optimal way\footnote{The max-weight policy \cite{kadota18} is effectively observing the same rule.}. Although several works have proposed optimization techniques, the explicit AoI analysis is relatively scarcely treated---an AoI lower bound is obtained in \cite{kadota18} which is conjectured to be asymptotically tight. 

Hybrid ARQ (HARQ) is widely-used in modern wireless systems, which differs from standard ARQ in the way previous transmissions are treated---HARQ effectively combine historical transmissions of same packets whereas standard ARQ discards them. At the expense of affordable additional complexity, the transmission reliability is significantly improved and hence HARQ is adopted in almost every wireless system in today's networks. The consideration of HARQ in the context of AoI presents distinct challenges. Unlike conventional packet transmissions wherein only reliability matters, optimizing AoI forces the HARQ mechanism to consider whether it is worthwhile to repeat an old packet---the tradeoff lies in that repetition is definitely more likely to succeed but sacrificing timeliness. In \cite{ceran19}, it is found that the optimal policy to minimize AoI with HARQ follows a threshold-based manner, for a single-link scenario.

In this paper, we are particularly interested in the RR-P policy. It was shown in \cite{jiang18_iot} that round-robin scheduling is asymptotically optimal when stochastic arrivals and reliable channels are considered. However, as mentioned previously, RR-P in scenarios with heterogeneous terminals' channels renders arbitrarily worse performance compared with optimum, due to the fact that, intuitively, a terminal with a very bad channel would jam the system for a long time. Despite of this, RR-P has several practically desirable merits, and surprisingly, it will be shown that its performance is in fact very close to optimum when HARQ is considered. Major advantages of RR-P include:

	\textbf{Channel-agnostic:} Unlike Whittle's index policy which needs to be aware of the channel conditions of all terminals, effectively entailing a pilot overhead which is often omitted in existing works, RR-P does not need any channel knowledge.

	\textbf{Easy-to-decentralize:} A round-robin-type scheduling is friendly to decentralized access. A token ring passing channel access scheme, e.g., in IEEE 802.5 \cite{token}, can be utilized to realize round-robin scheduling. Distributed terminals only need to know their local AoI, which is in fact easy by a simple acknowledgment feedback. In contrast, index-based policies need to compare all terminals' states, hindering decentralized implementation.

	\textbf{Near-optimality:} Even with its simplicity, RR-P is proved to be very close to optimum with a large number of terminals. Specifically, we can prove rigorously that under practical HARQ models, the asymptotic AoI loss, in terms of relative AoI increase compared with optimum with many terminals, is within a constant of $6.4\%$ in a typical scenario. In practice, the real AoI loss is even smaller exhibited by computer simulations.
	
In Section \ref{sec_model}, we will introduce the system model under consideration, including HARQ models and AoI evolution; the main results are presented in Section \ref{sec_opt}, wherein we derive theoretical AoI lower bound and achievable AoI by RR-P, and further show that they are close. Simulation results to numerically exhibit the performance is given in Section \ref{sec_sr}. Several proof details are presented in the Appendix.

\section{System Model}
\label{sec_model}
We consider a one-hop wireless network wherein a central node communications with $N$ distributed terminals. The terminals share the wireless channel based on a scheduling policy denoted by $\boldsymbol{\pi}$. A time-slotted status update system is considered. The status packet generation is assumed to be generate-at-will, i.e., a fresh status for terminal-$n$ is generated whenever it is scheduled. We are interested in average AoI. Concretely, the $T$-horizon time-average AoI of the system is defined by
\begin{equation}
\label{AoI}
\Delta_{\boldsymbol{\pi}}^{(T)} \triangleq \frac{1}{T N}\sum_{t=1}^T \sum_{n=1}^N \mathbb{E}[h_{n,\boldsymbol{\pi}}(t)],
\end{equation}
where $T$ is the time horizon length, and $h_{n,\boldsymbol{\pi}}(t)$ denotes the AoI of terminal-$n$ at the $t$-th time slot under policy $\boldsymbol{\pi}$. The long-time average AoI if defined by
\begin{equation}
\label{aoi_inf}
\bar{\Delta}_{\boldsymbol{\pi}} \triangleq \limsup_{T \to \infty} \Delta_{\boldsymbol{\pi}}^{(T)}.
\end{equation}

\subsection{Status Updates with HARQ}
We assume a perfect (i.e., error- and delay-free) one-bit feedback channel from the status update destination to the source node. In case of a successful reception of a status update packet, the destination feeds back an ACK; otherwise a NACK is fed back to indicate a transmission failure. In principle, retransmissions based on the feedback have the potential to improve the performance. Therefore, HARQ is considered in this paper. There are many different HARQ schemes in the literature. As a convention, they are categorized into two types. First, the type-I HARQ schemes, by which the destination node discards previous transmitted packets and treats each (re)transmissions are new---this is similar with standard ARQ except for the naming convention. Secondly, the type-II HARQ schemes combine (re)transmissions of the same packet for lower packet error performance, at the expense of more complicated buffer and algorithm design. Furthermore, there two widely-used type-II HARQ schemes:
\begin{itemize}
	\item 
	Chase Combining HARQ (CC-HARQ): The receiver uses Maximum Ratio Combining (MRC) to achieve a signal power gain, and all (re)transmissions carry the same coded bits. The MRC is implemented on the symbol level before the channel decoder.
	\item
	Incremental Redundancy HARQ (IR-HARQ): The information bits are coded with incremental redundant bits for error correction, each increment is carried in a retransmission. The receiver combines the coded bits of (re)transmissions and feeds into the channel decoder. 
\end{itemize}

One distinct tradeoff for type-II HARQ in status update is between the transmission success probability and the status freshness, in light of the fact that retransmissions carry the same old information dated back to the original transmission. Whereas type-I HARQ discards old packets anyway, it can always transmit fresh information. Without going into much details about HARQ which is out of the scope of this paper, we consider two models of packet error probability, i.e.,
\begin{equation}
\label{gr}
\operatorname{g}_{n,1}(r) = \frac{p_{n,0}}{r+1},\,\operatorname{g}_{n,2}(r) = p_{n,0} \lambda^{r},
\end{equation}
where $\operatorname{g}_{n,i}(r)$ denotes the packet error probability after the $r$-th (re)transmissions, $r\in\{0,1,2,...\}$. The packet error probability of the first transmission (or type-I HARQ retransmissions) for terminal-$n$ is denoted by $p_{n,0} \in [0,1]$, which can be different among terminals, and $\lambda \in (0,1)$ is a parameter related to HARQ protocol and channel conditions. It is noted that $\operatorname{g}_1(r)$ is suited for an i.i.d. fading scenarios with sufficient coding blocklength \cite{chai13}, whereas $\operatorname{g}_2(r)$ is more appropriate to model finite blocklength effects in quasi-static channels \cite{ceran19}. A detailed justification is presented in Appendix \ref{app_harq}. We further assume that the packet lengths and transmit power of (re)transmissions are the same, and the maximum number of retransmissions is unlimited. Each packet transmission is a independent Bernoulli trail with fail probability given above. The following lemma is useful in our analysis, regarding the average consecutive transmission attempts for a successful delivery. 

\begin{lemma}
	\label{lmInterval}
	With Eq. \eqref{gr}, the first and second moments of the number of consecutive transmission attempts for a successful delivery satisfy
	\begin{iarray}
	\mathbb{E}[K_1] &\triangleq& \sum_{r=0}^{+\infty}  \left[\prod_{i=0}^{r-1}\operatorname{g}_1(i) (1-\operatorname{g}_1(r)) (r+1)\right]= e^{p_0}, \nonumber\\
	\mathbb{E}[K_1^2] &\triangleq& \sum_{r=0}^{+\infty}  \left[\prod_{i=0}^{r-1}\operatorname{g}_1(i) (1-\operatorname{g}_1(r)) (r+1)^2\right]= (1+2p_0)e^{p_0}, \nonumber\\
	\mathbb{E}[K_2] &\triangleq& \sum_{r=0}^{+\infty}  \left[\prod_{i=0}^{r-1}\operatorname{g}_2(i) (1-\operatorname{g}_2(r)) (r+1)\right] \nonumber\\
	&\le& 1 + \left(1+\sqrt{\frac{2\pi}{-\log{\lambda}}}\right)p_0, \nonumber\\
	\mathbb{E}[K_2^2] &\triangleq& \sum_{r=0}^{+\infty}  \left[\prod_{i=0}^{r-1}\operatorname{g}_2(i) (1-\operatorname{g}_2(r)) (r+1)^2\right] \nonumber\\
	&\le& \frac{2\log{p_0}-2}{\log{\lambda}}-1+\left(2-\frac{2\log{p_0}}{\log{\lambda}}\right)\mathbb{E}[K_2], 
	\end{iarray}
	respectively, where the terminal index is omitted, and we prescribe $\operatorname{g}_{i}(-1)=1$, $i=1,2$.
\end{lemma}
\begin{IEEEproof}
	See Appendix \ref{app_interval}.
\end{IEEEproof}

\subsection{State, Action and Problem Formulation}
At each time slot, the state of terminal-$n$ is defined as $s_{n}(t) \triangleq (h_{n}(t),\,r_{n}(t))$, wherein $r_{n}(t)$ denotes the number of previous (re)transmissions of the same packet. Note that a reasonable policy would not re-send an older packet, since the policy has decided to transmit a new packet in previous time slots. 

The scheduling action includes deciding which terminal to be scheduled, and whether it should re-transmit, if any, an old packet, or transmit a new one. Formally, the action space is denoted by
$\mathcal{A} \triangleq \{n_{\mathsf{x}}|n\in\{1,...,N\},\,\mathsf{x}\in\{\mathsf{n},\mathsf{o}\}\}\}$, wherein $\mathsf{x}=\mathsf{n}$ and $\mathsf{x}=\mathsf{o}$ denote transmitting a new packet and re-transmitting an old one, respectively. The state transition probability is hence written as
\begin{iarray}
	&& \Pr\{ h_n+1,1|h_{n},r_n,n_{\mathsf{n}}\} = \operatorname{g}(0); \nonumber\\
	&& \Pr\{ 1,0|h_{n},r_n,n_{\mathsf{n}}\} = 1-\operatorname{g}(0); \nonumber\\
	&& \Pr\{ h_n+1,r_n+1|h_{n},r_n,n_{\mathsf{o}}\} = \operatorname{g}(r); \nonumber\\
	&& \Pr\{ r_n+1,0|h_{n},r_n,n_{\mathsf{o}}\} = 1-\operatorname{g}(r); 
\end{iarray}
and when terminal-$n$ is not scheduled,
\begin{iarray}
	&& \Pr\{ h_n+1,r_n|h_{n},r_n,i_{\mathsf{x}},i \neq n\}=1, 
\end{iarray}
and other transition probabilities equal zero.

We assume that in each time slot, only one terminal can be scheduled. The objective is to find a policy $\boldsymbol{\pi}$ that minimizes the long-term average AoI in \eqref{aoi_inf}, and to analyze its performance. In most part of the paper, we consider a large number of terminals, i.e., $N \to \infty$.

\section{Optimality of RR-P with Type-II HARQ}
\label{sec_opt}
In this section, the asymptotic optimality of RR-P when the number of terminals $N$ is large is shown. The method is based on first finding an AoI lower bound which leverages a similar method in \cite{kadota18}, and then deriving an achievable AoI analytical results (upper bound) by the RR-P. By showing that the gap in between is vanishing, it can be concluded that RR-P is asymptotically optimal.
\subsection{AoI Lower Bound with HARQ}
The AoI is shown to have the following property \cite[Lemma 1]{ery13}, \cite[Theorem 6]{kadota18}.
\begin{lemma}
	For a scheduling policy that schedules every terminal infinitely often, i.e., ergodic, the long-time average AoI satisfies
	\begin{equation}
	\bar {\Delta}_{\boldsymbol{\pi}} \ge \frac{1}{2N} \sum_{n=1}^{N} \frac{\bar{\mathbb{M}} [\delta_n^2]}{\bar{\mathbb{M}} [\delta_n]} + \frac{1}{2},
	\end{equation}
	where $\bar{\mathbb{M}}(\cdot)$ denotes the sample mean, and $\delta_n$ is the inter-delivery time of terminal-$n$, i.e., number of time slots between consecutive successful deliveries. In addition, if the policy is also renewal,  
	\begin{equation}
	\bar {\Delta}_{\boldsymbol{\pi_\mathsf{R}}} \ge \frac{1}{2N} \sum_{n=1}^{N} \frac{{\mathbb{E}} [\delta_n^2]}{{\mathbb{E}} [\delta_n]} + \frac{1}{2},
	\end{equation}
	where $\bar{\mathbb{E}}(\cdot)$ denotes the expectation.
\end{lemma}
\begin{IEEEproof}
The proposition is satisfied with equality in previous works without considering HARQ, i.e., the corresponding AoI is reset to $1$ after each successful delivery. Therefore, the AoI with HARQ is lower bounded by the expressions in the lemma, considering that the AoI would be reduced to the time duration since the first transmission of the current packet which is larger or equal to one (in case of this is the first attempt, the AoI would return to one).
\end{IEEEproof}

This Lemma clearly shows the relationship between inter-delivery time and AoI, and is leveraged in the rest of the paper. Note that the condition of ergodicity is not restrictive, since a policy that starves any terminal is apparently sub-optimal. Also note that a renewal policy is defined as one that results in i.i.d. inter-delivery time. Hence a stationary policy that schedules any terminal based on a time-invariant probability is included; the RR-type policy is also included based on the definition.

\begin{lemma}[Lower Bound]
	\label{lmLb}
	The long time-average AoI is lower bounded by
	\begin{equation}
	\bar{\Delta}_{\boldsymbol{\pi}} \ge \frac{1}{2N} \left(\sum_{n=1}^N \sqrt{\mathbb{E}[K_{i,n}]}\right)+\frac{1}{2}
	\end{equation}
\end{lemma}
\begin{IEEEproof}
	Denote the number of successful deliveries of terminal-$n$ up to the $L$-th time slot as $D_n(L)$, and the number of transmission attempts of terminal-$n$ up to the $L$-th time slot as $A_n(L)$, then
	\begin{iarray}
	\bar{\Delta}_{\boldsymbol{\pi}} &\overset{(a)}{\ge}& \frac{1}{2N} \sum_{n=1}^N \bar{\mathbb{M}} [\delta_n]+\frac{1}{2} \nonumber\\
	&\overset{(b)}{=}&  \frac{1}{2N} \sum_{n=1}^N \lim_{L \to \infty} \frac{L}{D_n(L)}+\frac{1}{2}	 \nonumber\\
	&\overset{(c)}{\ge}&  \frac{1}{2N} \lim_{L \to \infty} \sum_{n=1}^N A_n(L) \sum_{m=1}^N \frac{1} {D_m(L)}+\frac{1}{2} \nonumber\\
	&\overset{(d)}{\ge}&  \frac{1}{2N} \left(\sum_{n=1}^N \sqrt{\lim_{L \to \infty} \frac{A_n(L)} {D_n(L)}}\right)^2+\frac{1}{2} \nonumber\\
	&\overset{(e)}{\ge}& \frac{1}{2N} \left(\sum_{n=1}^N \sqrt{\mathbb{E}[K_{i,n}]}\right)^2+\frac{1}{2} \nonumber\\
	&=& \frac{N}{2} \bar{\mathbb{M}} \left[ \sqrt{g(\boldsymbol{\omega}_{i,n})}\right]^2+\frac{1}{2}, 
	\end{iarray}
which concludes the proof. Denote by $\bar{\mathbb{M}} \left[ \sqrt{\mathbb{E}[K_{i,n}]}\right] \triangleq \frac{\sum \sqrt{\mathbb{E}[K_{i,n}]}}{N}$ as the sample mean among the terminals, and $g(\boldsymbol{\omega}_{i,n}) \triangleq \mathbb{E}[K_{i,n}]$ is a function of channel parameters, i.e., $\boldsymbol{\omega}_{1,n}=[p_{n,0}]$ and $\boldsymbol{\omega}_{2,n}=[p_{n,0},\lambda]$. The inequality $(a)$ follows from $\bar{\mathbb{M}} [\delta_n^2] \ge \bar{\mathbb{M}} [\delta_n]^2$, wherein the equality holds when the variance is zero. The equality $(b)$ is obtained by definition. The inequality $(c)$ is because $L\ge\sum_{n=1}^N A_n(L)$ since there are altogether $L$ time slots. The Cauchy-Schwarz inequality gives $(d)$, and the last inequality $(e)$ follows from the fact that the minimum average transmission attempts required to reach a successful delivery by HARQ is obtained by successive repetitive transmissions of old packets. The average is given by Lemma \ref{lmInterval}.
\end{IEEEproof}

\begin{remark}
When type-I HARQ is considered, i.e., equivalent with standard ARQ in this context, the lower bound results in \cite{kadota18} is a special case of this lemma, wherein $\frac{A_n(L)} {D_n(L)}$ tends to the inverse of the transmission error probability. 
\end{remark}

Denote the AoI lower bound in Lemma \ref{lmLb} as $\bar{\Delta}_{\mathsf{LB}}$, then the following corollary follows straightforwardly. 
\begin{coro}
\begin{equation}
\label{gap}
\bar{\Delta}_{\mathsf{LB}} \le \frac{N}{2} \bar{\mathbb{M}} \left[ g(\boldsymbol{\omega}_{i,n})\right]+\frac{1}{2}
\end{equation}
\end{coro}
\begin{IEEEproof}
The inequality follows from $\bar{\mathbb{M}} [x]^2 \le \bar{\mathbb{M}}[x^2]$. In the subsequent section, we will see that this corollary reflects the gap between the lower bound and the achievable AoI by the RR-P policy, as RR-P achieves (approximately) the RHS of \eqref{gap}.
\end{IEEEproof}
\subsection{Achievable AoI by RR-P}
\begin{defi}
	The RR-P scheduling policy schedules the terminals in a round-robin manner. When scheduled, the terminal will transmit and re-transmit the same packet until successful delivery. 
\end{defi}
The achievable AoI by RR-P is shown by the following theorem. 
\begin{theorem}
	\label{thmRR}
	Under the HARQ models in \eqref{gr}, the long time-average AoI achieved by RR-P is 
	\begin{equation}
	\label{rrpExact}
	\bar{\Delta}_{\mathsf{RR\_ P},i} =  \bar{\mathbb{M}} \left[ g(\boldsymbol{\omega}_{i,n})\right] +\frac{1}{2} \frac{\mathbb{E}\left[\left(\sum_{n=1}^N [K_{i,n}]\right)^2\right]}{\bar{\mathbb{M}} \left[ g(\boldsymbol{\omega}_{i,n})\right]} - \frac{1}{2},
	\end{equation}
which satisfies
\begin{equation}
\label{rrpBound}
\bar{\Delta}_{\mathsf{RR\_P},i} \le \frac{N+1}{2} \bar{\mathbb{M}} \left[ g(\boldsymbol{\omega}_{i,n})\right] +\frac{1}{2} \frac{\bar{\mathbb{M}} \left[ \mathbb{E}[K_{i,n}^2]\right]}{\bar{\mathbb{M}} \left[ g(\boldsymbol{\omega}_{i,n})\right]} - \frac{1}{2}.
\end{equation}
Furthermore, 
\begin{equation}
\label{rrpAsym}
\frac{N}{2} \bar{\mathbb{M}} \left[ g(\boldsymbol{\omega}_{i,n})\right] -\frac{1}{2} \le \bar{\Delta}_{\mathsf{RR\_P},i} \le \frac{N}{2} \bar{\mathbb{M}} \left[ g(\boldsymbol{\omega}_{i,n})\right] + c,
\end{equation}
where $c$ is a constant irrelevant with $N$. This inequality gives the asymptotic scaling factor when the number of terminals is large, i.e.,
\begin{equation}
\lim_{N\to\infty} \frac{\bar{\Delta}_{\mathsf{RR\_P},i}}{N} = \frac{\bar{\mathbb{M}} \left[ g(\boldsymbol{\omega}_{i,n})\right]}{2} 
\end{equation}
\end{theorem}
\begin{IEEEproof}
See Appendix \ref{app_thm}.	
\end{IEEEproof}

Armed with this theorem, in particular the asymptotic results, the relative AoI gap between RR-P and the optimum (i.e., $(1+\gamma)$-optimality where $\gamma$ denotes the relative gap) with a large number of terminals can be studied. In the following subsection, explicit and tight results will be presented.
\subsection{Asymptotic $(1+\gamma)$-Optimality of RR-P}
We investigate the asymptotic order-optimality of RR-P, that is, the relative AoI gap compared with optimum when the number of terminals is large. Define the asymptotic relative AoI gap of RR-P as
\begin{equation}
\gamma_i \triangleq \lim_{N\to\infty} \frac{\bar{\Delta}_{\mathsf{RR\_P},i}-\bar{\Delta}_{\mathsf{opt}}}{\bar{\Delta}_{\mathsf{opt}}}.
\end{equation}
Based on Lemma \ref{lmLb} and Theorem \ref{thmRR}, the gap is smaller or equal to
\begin{equation}
\label{gapBound}
\gamma_i \le \lim_{N\to\infty} \frac{\bar{\Delta}_{\mathsf{RR\_P},i}-\bar{\Delta}_{\mathsf{LB}} }{\bar{\Delta}_{\mathsf{LB}}}=\frac{\bar{\mathbb{M}} \left[ g(\boldsymbol{\omega}_{i,n})\right]-\bar{\mathbb{M}} \left[ \sqrt{g(\boldsymbol{\omega}_{i,n})}\right]^2}{\bar{\mathbb{M}} \left[ \sqrt{g(\boldsymbol{\omega}_{i,n})}\right]^2}.
\end{equation}
The following theorem explicitly bound the gap.
\begin{theorem}
	\label{thmGap}
Under the HARQ models in \eqref{gr} with relative gaps $\gamma_i$, $i \in \{1,2\}$, RR-P is within $(1+\gamma_i)$-optimality with $N \to \infty$, and 
\begin{iarray}
\gamma_1 &\le& \frac{(\sqrt{e}-1)^2}{4\sqrt{e}} \cong 6.4\%,\nonumber\\
\gamma_2 &\le& \frac{\left(\sqrt{2+\sqrt{\frac{2 \pi}{-\log{\lambda}}}}-1\right)^2}{4 \sqrt{2+\sqrt{\frac{2 \pi}{-\log{\lambda}}}}}.
\end{iarray}	
\end{theorem}
\begin{IEEEproof}
	Following \eqref{gapBound},
\begin{iarray}
	\gamma_i &\le& \frac{\bar{\mathbb{V}} \left[ \sqrt{g(\boldsymbol{\omega}_{i,n})}\right]}{\bar{\mathbb{M}} \left[ \sqrt{g(\boldsymbol{\omega}_{i,n})}\right]^2} \nonumber\\
	&\overset{(a)}{\le}& \frac{\left(\sqrt{g_{\mathsf{max},i}} - \bar{\mathbb{M}} \left[ \sqrt{g(\boldsymbol{\omega}_{i,n})}\right] \right)\left(\bar{\mathbb{M}} \left[ \sqrt{g(\boldsymbol{\omega}_{i,n})}\right] -  \sqrt{g_{\mathsf{min},i}}\right)}{\bar{\mathbb{M}} \left[ \sqrt{g(\boldsymbol{\omega}_{i,n})}\right]^2} \nonumber\\
	&\le& \frac{\left(\sqrt{g_{\mathsf{max},i}} -  \sqrt{g_{\mathsf{min},i}}\right)^2}{4\sqrt{g_{\mathsf{max},i} g_{\mathsf{min},i}}},\nonumber
\end{iarray}	
wherein $\bar{\mathbb{V}}[\cdot]$ denotes the variance operator, and the inequality $(a)$ stems from \cite{variance}. Based on Lemma \ref{lmInterval}, we select
\begin{equation}
g_{\mathsf{min},i} = 1,\, i=1,2.
\end{equation}
\begin{equation}
g_{\mathsf{max},1} = e, \, g_{\mathsf{max},2} = 2+\sqrt{\frac{2 \pi}{-\log{\lambda}}},
\end{equation}
and hence the conclusion follows immediately.
\end{IEEEproof}
\begin{remark}
	Based on Theorem \ref{thmGap}, it is shown that when the number of terminals is large, the relative AoI increase by RR-P compared with optimum is within $6.4$ percents with the first HARQ model; for a practical value of $\lambda=0.5$, the gap is within $17.1$ percents with the second HARQ model. Note that this does not mean the performance loss with the second model is larger---this is mainly due to the fact that we can only obtain an upper bound with $K_2$, which is often loose. In fact, applying a better bound by Corollary \ref{coroK2} with $R=4$, we can show the $\gamma_2 \le 6.2\%$ with $\lambda=0.5$. 
	
	In contrast, based on \cite[Theorem 8]{kadota18}, no constant $\gamma$ can be found for RR-P with type-I HARQ, or equivalently standard ARQ in this context. In other words, RR-P can be arbitrarily worse than optimum with standard ARQ (transmitting a new packet at each opportunity). In fact, as far as we know, the best proven bound for standard ARQ with unequal error probabilities is $\gamma=1$ (i.e., $2$-optimal), using a stationary randomized policy with optimized transmission probabilities. Policies with simulated better performance, e.g., Whittle index policy, can only be proven with very loose bounds which render meaningless due to their the non-renewal nature, resulting in difficulties in age analysis. The bounds in this paper are much tighter, in scenarios with HARQ which essentially favors retransmissions and hence makes the analysis more tractable.   
\end{remark}
\begin{figure*}[!t]
	\centering
	\subfigure[]{
		\includegraphics[width=0.45\textwidth]{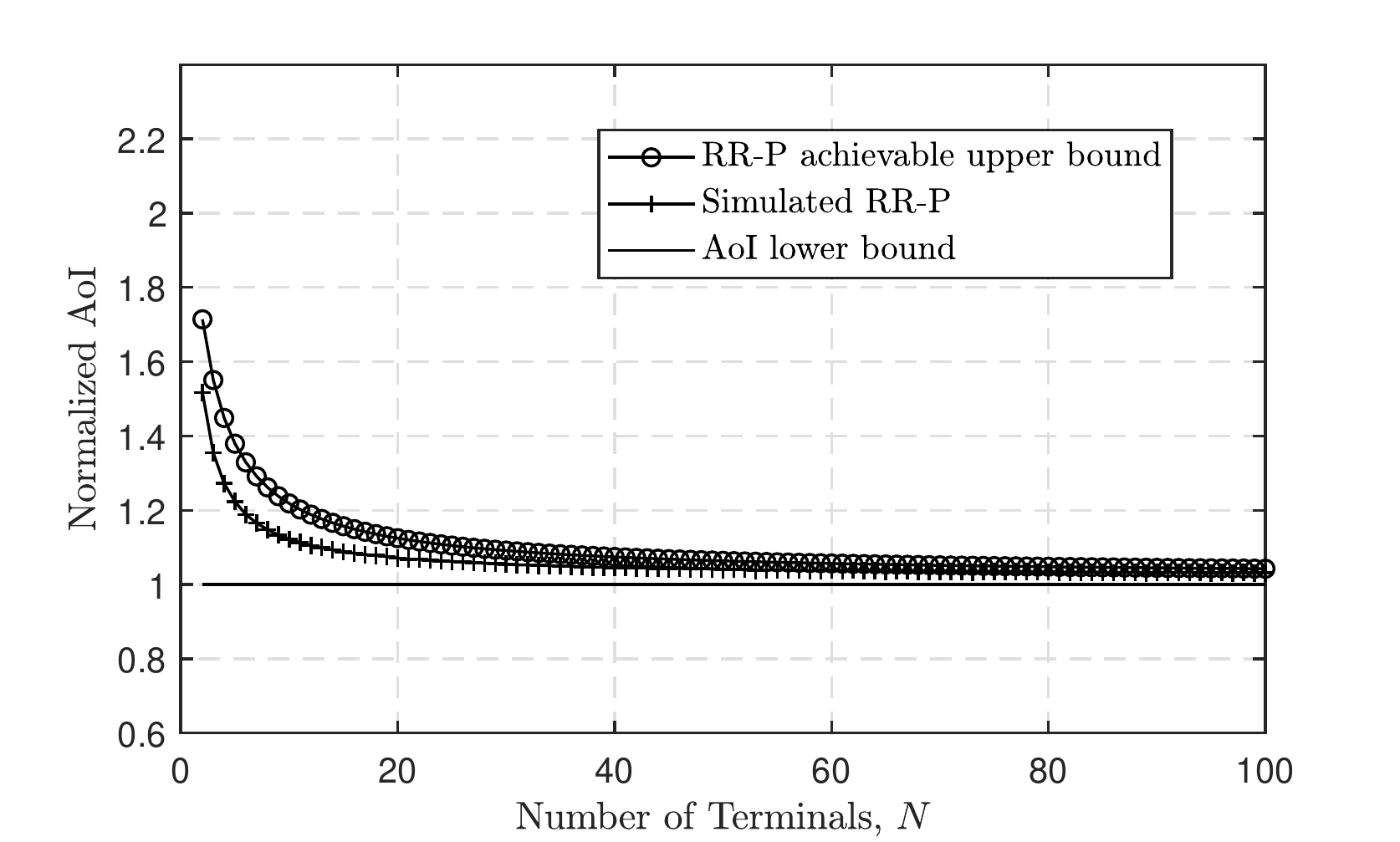}}
	\subfigure[]{
		\includegraphics[width=0.45\textwidth]{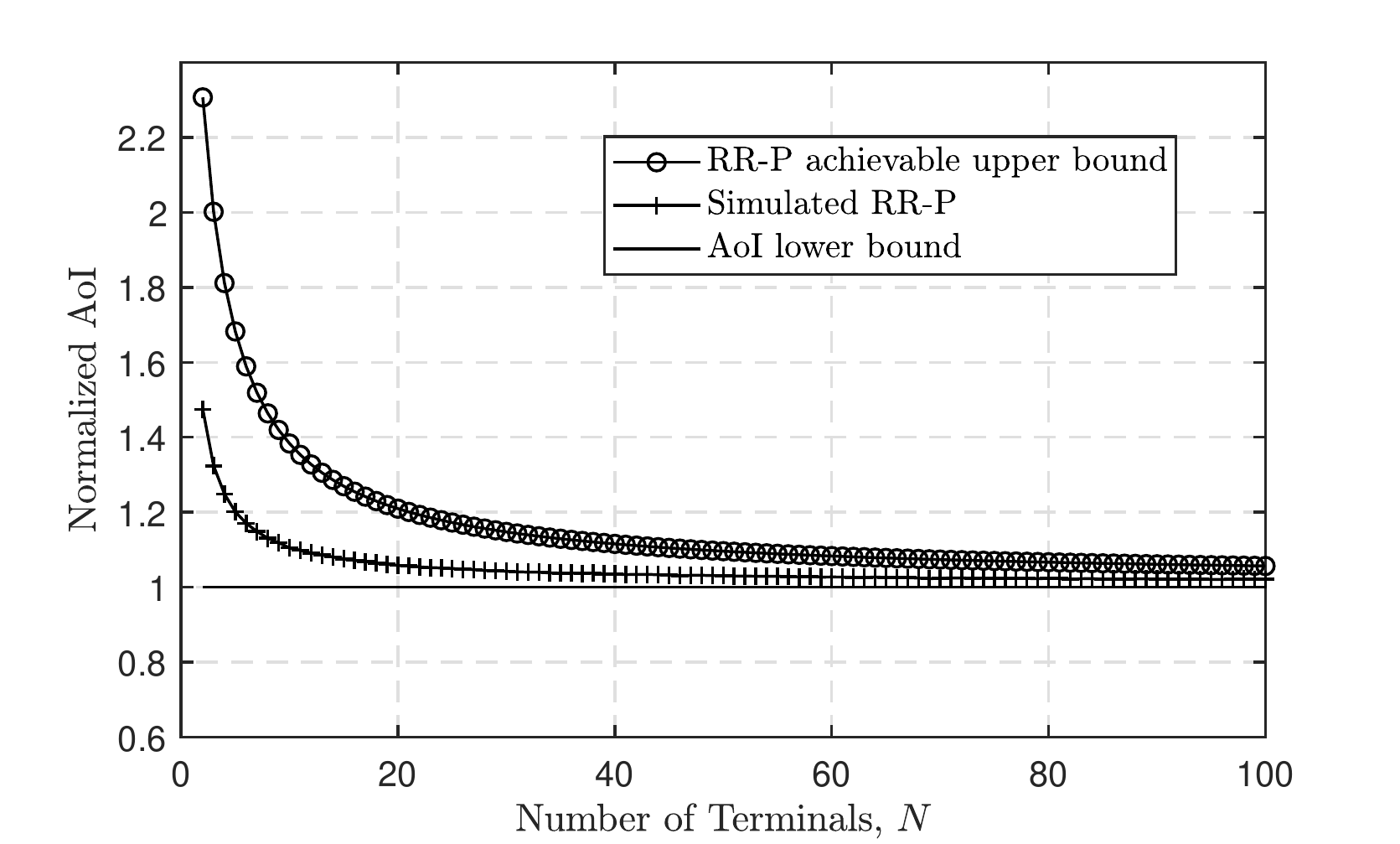}}
	\caption{AoI normalized by the lower bound in Lemma \ref{lmLb}, achieved by the RR-P through Monte-Carlo simulations and its upper bound in Theorem \ref{thmRR}. HARQ models in \eqref{gr} are both shown in subfigure (a) and (b), respectively. $\lambda=0.5$ in subfigure (b), and the upper bound is calculated based on Corollary \ref{coroK2} with $R=4$.}
	\label{fig_harqModel}
\end{figure*}
\section{Simulation Results}
\label{sec_sr}
We run computer simulations for RR-P for $10^6$ time slots and take the time-average AoI. The AoI lower bound in Lemma \ref{lmLb} and achievable AoI upper bound by RR-P in Theorem \ref{thmRR} are calculated and compared in the figures. The number of terminals $N$ varies from $3$ to $100$. The corresponding initial transmission error probabilities $p_{n,0}$, $n=1,\cdots,N$ are set to $[1/N,2/N,\cdots,1]$, respectively. The AoI is normalized by the lower bound derived by Lemma \ref{lmLb}. It is observed that, with both HARQ models, the relative AoI increase by RR-P approaches very small with the number of terminals greater than, e.g., $20$. The upper bound that is used to prove the main result follows this trend closely, and moreover, the actual RR-P performance obtained by Monte-Carlo simulations is even closer to the lower bound, indicating that RR-P can performance even better than the proved theoretical results in practice. 

\section{Conclusions}
\label{sec_cc}
This paper has shown that RR-P is provably near-optimal for AoI optimizations with HARQ in heterogeneous unreliable multiaccess channels wherein terminals have distinctive transmission error probabilities and the number of terminals is large. Concretely, it is proved that the relative AoI gap by RR-P compared with the optimum is within a constant of $(\sqrt{e}-1)^2/4\sqrt{e} \cong 6.4\%$ (resp. $6.2\%$ with error exponential decay rate of $0.5$) with fading channels (resp. finite blocklength scenarios) asymptotically. In reality, the gap becomes even smaller than the theoretical bounds, which is shown by computer simulations. The simulation results also reveal that the number of terminals required for the asymptotic results to hold is approximately $20$. Moreover, the gap increases with the terminals transmission error heterogeneity, i.e., the variance of terminals' transmission error probabilities.

The results in this paper relies crucially on the renewal structure of RR-P. It is still difficult to obtain closed-form AoI analysis for non-renewal policies, as evidenced by several studies in the literature \cite{kadota18,ali20}. More advanced mathematical tools are needed to address this issue in future works.

\appendices
\section{Justification of HARQ Models of \eqref{gr}}
\label{app_harq}
Without loss of generality, let us consider one representative terminal. Assuming a block-fading Rayleigh channel based on which the complex baseband channel stays constant during each HARQ transmission round and changes to another value based on an i.i.d. complex Gaussian distribution. The block error probability in each round is approximated by the information outage probability which is defined to be the probability that instantaneous spectral efficiency given by Shannon formula is smaller than the target spectral efficiency. Furthermore, assume that the transmission power stays the same and CC-HARQ is adopted. Then the block error rate in the $r$-th round can be well approximated by the first model in \eqref{gr} based on \cite[Theorem 1]{chai13}. More precisely, the probability that the first $l$ transmissions all fail is approximately
\begin{equation}
p_{\mathsf{out},l} \cong \frac{p_{\mathsf{out},1}^{l}}{l!} + \mathcal{O}(p_{\mathsf{out},1}^{l+1}),
\end{equation}
wherein the factorial term represents the power gain by CC-HARQ.

On the other hand, consider the finite blocklength regime and a non-fading AWGN channel, wherein the block error, instead of fading,  is mainly caused by insufficient channel coding bits and white noise. For ease of exposition, consider a Binary Erasure Channel (BEC) for each bit with erasure rate of $\delta$. Consider IR-HARQ, a message can only be correctly decoded when the total number of successful bits is more than $C$, and the total number of transmitted bits is $lB$ where $B$ is the blocklength of one transmission and $l$ is total transmission rounds. Therefore, it follows that the error probability follows the cumulative distribution function of binomial distribution, i.e.,
\begin{equation}
p_{\mathsf{e},l} = \sum_{c=0}^{C-1} { lB \choose c}(1-\delta)^c\delta^{lB-c}.
\end{equation}
It is well-known that when $lB$ is large enough compared with $C$,\footnote{A common condition is that \cite{box05} $\frac{|1-2\delta|}{\sqrt{lB\delta(1-\delta)}}<\frac13$.} a reasonable approximation of the binomial distribution is Gaussian distribution of $\mathcal{N}(lB\delta,\, lB\delta(1-\delta))$, i.e., the probability mass function can be approximated by 
\begin{equation}
f_{\mathsf{e},l,c} \cong \frac{1}{\sqrt {2\pi lB\delta(1-\delta)}}e^\frac{-(c-lB\delta)^2}{2lB\delta(1-\delta)}.
\end{equation}
Therefore, the success probability after $l$ rounds is approximated by
\begin{iarray}
p_{\mathsf{e},l} &\cong& \operatorname{Q}\left(\frac{\left|C-lB\delta\right|}{\sqrt{lB\delta(1-\delta)}}\right) \lessapprox e^{-\frac{\left(C-lB\delta\right)^2}{2lB\delta(1-\delta)}} \nonumber\\
&\overset{lB \gg C}{\cong}& e^{-\frac{lB\delta}{2(1-\delta)}},
\end{iarray}
which coincides with the second model in \eqref{gr} whereby the error probability scales down exponentially with the number of (re)transmission attempts. Note that by definition, $l=r+1$. A similar, in fact much stronger arguments can be made based on \cite{fbl}, wheres the present paper provides a more intuitive explanation.

It can be observed that the two models in \eqref{gr} suit i.i.d. Rayleigh fading with CC-HARQ and finite blocklength with IR-HARQ methods, respectively. 

\section{Proof of Lemma \ref{lmInterval}}
\label{app_interval}
Considering $K_1$, we obtain
\begin{iarray}
	\mathbb{E}[K_1] &\triangleq& \sum_{r=0}^{+\infty}  \left[\prod_{i=0}^{r-1}\operatorname{g}_1(i) (1-\operatorname{g}_1(r)) (r+1)\right] \nonumber \\
	&=& \sum_{r=0}^{+\infty} \frac{p_0^r}{r!}\left(1-\frac{p_0}{r+1}\right)(r+1) \nonumber\\
	&=& \sum_{r=0}^{+\infty} \frac{p_0^r}{r!}(r+1) - \sum_{r=0}^{+\infty} \frac{p_0^r}{r!}r \nonumber\\
	&=& \sum_{r=0}^{+\infty} \frac{p_0^r}{r!} = e^{p_0}.
\end{iarray}
\begin{iarray}
	\mathbb{E}[K_1^2] &\triangleq& \sum_{r=0}^{+\infty}  \left[\prod_{i=0}^{r-1}\operatorname{g}_1(i) (1-\operatorname{g}_1(r)) (r+1)^2\right] \nonumber \\
	&=& \sum_{r=0}^{+\infty} \frac{p_0^r}{r!}\left(1-\frac{p_0}{r+1}\right)(r+1)^2 \nonumber\\
	&=& \sum_{r=0}^{+\infty} \frac{p_0^r}{r!}(2r+1) \nonumber\\
	&=& (1+2p_0)e^{p_0}.
\end{iarray}
Similarly, with $K_2$,
\begin{iarray}
	\label{k2}
	\mathbb{E}[K_2] &\triangleq& \sum_{r=0}^{+\infty}  \left[\prod_{i=0}^{r-1}\operatorname{g}_2(i) (1-\operatorname{g}_2(r)) (r+1)\right] \nonumber \\
	&=& \sum_{r=0}^{+\infty} {p_0^r} \lambda^{\frac{r(r-1)}{2}}\left(1-p_0\lambda^r\right)(r+1) \nonumber\\
	&=& \sum_{r=0}^{+\infty} {p_0^r} \lambda^{\frac{r(r-1)}{2}}(r+1) - \sum_{r=0}^{+\infty} {p_0^r} \lambda^{\frac{r(r-1)}{2}}r \nonumber\\
	&=& \sum_{r=0}^{+\infty} {p_0^r} \lambda^{\frac{r(r-1)}{2}} = 1+p_0+\sum_{r=2}^{+\infty} {p_0^r} \lambda^{\frac{r(r-1)}{2}} \nonumber\\
	&\overset{(a)}{\le}& 1+p_0+\int_1^{+\infty}{p_0^x} \lambda^{\frac{x(x-1)}{2}} \mathsf{d}x,
\end{iarray}
where inequality $(a)$ is due to the fact that for a monotonically decreasing function $f(x)=p_0^x\lambda^{\frac{x(x-1)}{2}}$, $x\in[2,+\infty)$, 
\begin{equation}
\sum_{r=2}^{+\infty}f(r) \le \int_1^{+\infty} f(x)\mathsf{d}x.
\end{equation}
Denote $\alpha \triangleq -\frac{\log \lambda}{2}$ and $\beta \triangleq -\log p_0$, then following \eqref{k2},
\begin{iarray}
	\mathbb{E}[K_2] &{\le}& 1+p_0+e^{\frac{(\alpha-\beta)^2}{4a}} \int_{\frac{1}{2}+\frac{\beta}{2\alpha}}^{+\infty} e^{-\alpha x^2} \mathsf{d}x, \nonumber\\
	&=& 1+p_0+e^{\frac{(\alpha-\beta)^2}{4a}} \sqrt{\frac{\pi}{\alpha}} \operatorname{Q}\left(\frac{\alpha+\beta}{\sqrt{2\alpha}}\right)\nonumber\\
	&\overset{(a)}{\le}& 1+p_0+\sqrt{\frac{\pi}{\alpha}} e^{-\beta} \nonumber\\
	&=& 1 + \left(1+\sqrt{\frac{2\pi}{-\log{\lambda}}}\right)p_0,
\end{iarray}
where $\operatorname{Q}(x) \triangleq \frac{1}{\sqrt{2\pi}}\int_x^{+\infty} e^{-t^2/2} \mathsf{d}t$ is the Q-function, and inequality $(a)$ follows from the Chernoff bound $\operatorname{Q}(x) \le e^{-x^2/2}$. The following corollary gives a tighter bound.
\begin{coro}
	\label{coroK2}
	\begin{equation}
		\mathbb{E}[K_2] {\le} \sum_{r=0}^{R-1} {p_0^r} \lambda^{\frac{r(r-1)}{2}} + \left(1+\sqrt{\frac{2\pi}{-\log{\lambda}}}\right) {p_0^R} \lambda^{\frac{R(R-1)}{2}},
	\end{equation}
wherein $R \in \mathbb{N}^+$.
\end{coro}
\begin{IEEEproof}
\begin{iarray}
	\mathbb{E}[K_2] &{\le}& \sum_{r=0}^{R} {p_0^r} \lambda^{\frac{r(r-1)}{2}} +\sum_{r=R+1}^{+\infty} {p_0^r} \lambda^{\frac{r(r-1)}{2}} \nonumber\\
	&\le& \sum_{r=0}^{R} {p_0^r} \lambda^{\frac{r(r-1)}{2}} + \int_{R}^{+\infty}{p_0^x} \lambda^{\frac{x(x-1)}{2}} \mathsf{d}x, \nonumber\\
	&=& \sum_{r=0}^{R} {p_0^r} \lambda^{\frac{r(r-1)}{2}} + e^{\frac{(\alpha-\beta)^2}{4a}} \sqrt{\frac{\pi}{\alpha}} \operatorname{Q}\left(\frac{(2R-1)\alpha+\beta}{\sqrt{2\alpha}}\right)\nonumber\\
	&\le& \sum_{r=0}^{R} {p_0^r} \lambda^{\frac{r(r-1)}{2}} + e^{\frac{(\alpha-\beta)^2-((2R-1)\alpha+\beta)^2}{4a}} \sqrt{\frac{\pi}{\alpha}} \nonumber\\
	&=& \sum_{r=0}^{R} {p_0^r} \lambda^{\frac{r(r-1)}{2}} + e^{-(R(R-1)\alpha+R\beta)} \sqrt{\frac{\pi}{\alpha}}, 
\end{iarray}
which concludes the proof.
\end{IEEEproof}
\begin{iarray}
	\mathbb{E}[K_2^2] &\triangleq& \sum_{r=0}^{+\infty}  \left[\prod_{i=0}^{r-1}\operatorname{g}_2(i) (1-\operatorname{g}_2(r)) (r+1)^2\right] \nonumber \\
	&=& \sum_{r=0}^{+\infty} {p_0^r} \lambda^{\frac{r(r-1)}{2}}\left(1-p_0\lambda^r\right)(r+1)^2 \nonumber\\
	&=& \sum_{r=0}^{+\infty} {p_0^r} \lambda^{\frac{r(r-1)}{2}}(r+1)^2 - \sum_{r=0}^{+\infty} {p_0^r} \lambda^{\frac{r(r-1)}{2}}r^2 \nonumber\\
	&=& \sum_{r=0}^{+\infty} {p_0^r} \lambda^{\frac{r(r-1)}{2}}(2r+1) \nonumber\\
	&=& \sum_{r=0}^{+\infty} {p_0^r} \lambda^{\frac{r(r-1)}{2}}2\left(r-\frac{\alpha-\beta}{2\alpha}\right)+\left(2-\frac{\beta}{\alpha}\right)\mathbb{E}[K_2] \nonumber\\
	&=& \sum_{r=1}^{+\infty} {p_0^r} \lambda^{\frac{r(r-1)}{2}}2\left(r-\frac{\alpha-\beta}{2\alpha}\right)\nonumber\\
	&& +\left(2-\frac{\beta}{\alpha}\right)\mathbb{E}[K_2]+\frac{\beta}{\alpha}-1 \nonumber\\
	&\le& e^{\frac{(\alpha-\beta)^2}{4\alpha}} \int_0^{+\infty}{e^{-\alpha \left(x-\frac{\alpha-\beta}{2\alpha}\right)^2}} \left(x-\frac{\alpha-\beta}{2\alpha}\right) \mathsf{d}x \nonumber\\
	&& +\left(2-\frac{\beta}{\alpha}\right)\mathbb{E}[K_2]+\frac{\beta}{\alpha}-1 \nonumber\\
	&=& \left(2-\frac{\beta}{\alpha}\right)\mathbb{E}[K_2]+\frac{\beta+1}{\alpha}-1,
\end{iarray}
which concludes the proof.
\section{Proof of Theorem \ref{thmRR}}
\label{app_thm}
\begin{figure}[!t]
\centering
\includegraphics[width=0.3\textwidth]{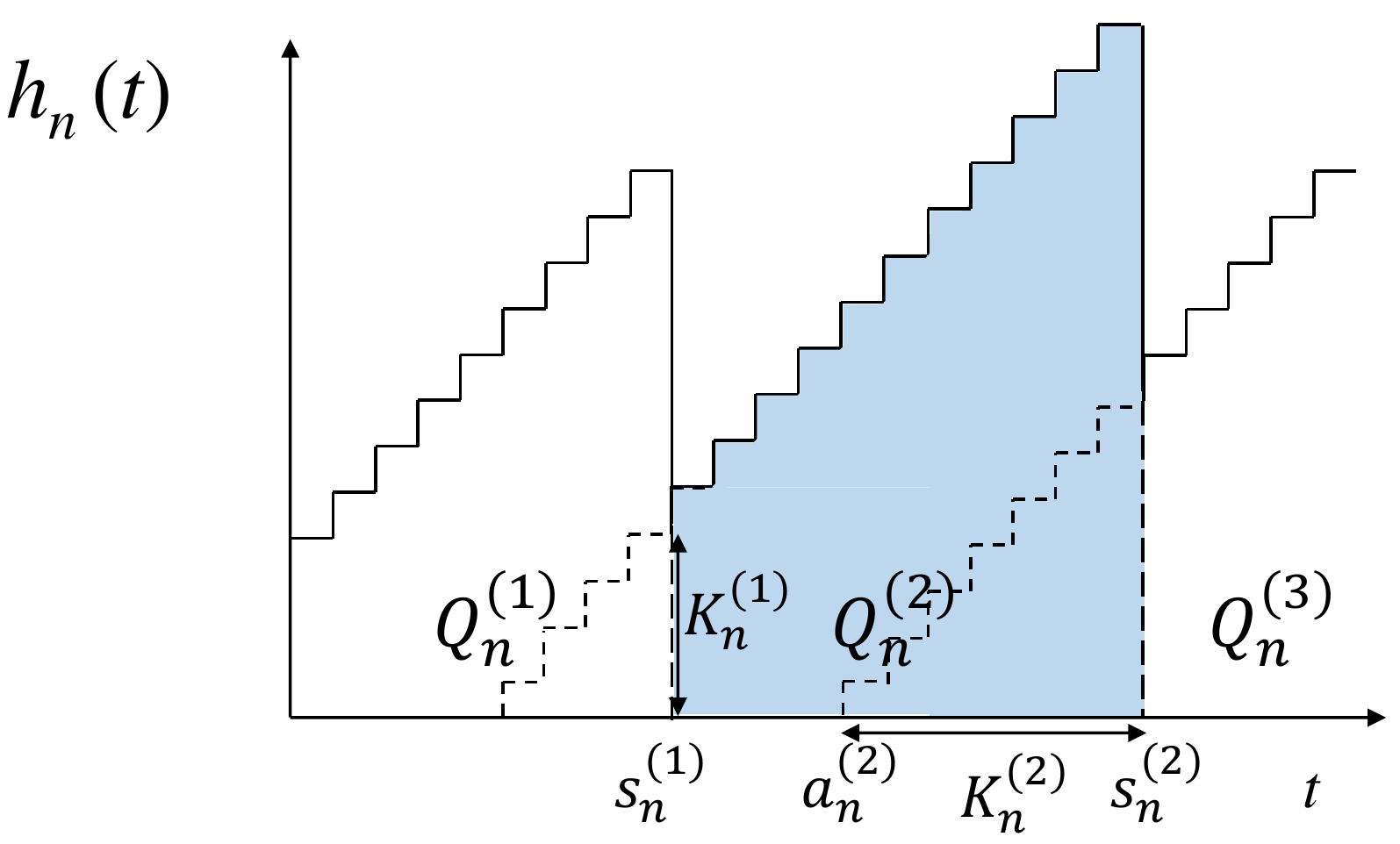}
\caption{AoI evolution of terminal-$n$ under HARQ.}
\label{fig_aoiHarq}
\end{figure}
Note that RR-P is a renewal policy that for each terminal, the $j$-th successful delivery interval is 
\begin{equation}
S_{n,\mathsf{RR\_P}}^{(j)} \triangleq \sum_{m=1}^N K_{m}^{(j)},\,\forall n \in \{1,...,N\},
\end{equation}
and $\{K_{m}^{(j)}|j =1,2,...\}$ are i.i.d. This is because based on RR-P, the successful delivery interval of every terminal is the total time that all terminals delivers an update. Therefore, without loss of generality, a sample path of the AoI evolution of terminal-$n$ is shown in Fig. \ref{fig_aoiHarq}. The time between the $i-1$-th and $i$-th deliveries is denoted by the $i$-th round, during which, the moment that the terminal is scheduled and transmits its first packet is denoted by $a_n^{(i)}$ in Fig. \ref{fig_aoiHarq}. The retransmissions continue until a successful delivery based on RR-P, and hence the age $h_n(t)$ drops to $K_{n}^{(i)}$ upon that---the time of which is denoted by $s_n^{(i)}$.  

Following the same arguments in, e.g., \cite{najm17}, the time-average AoI can be readily calculated by the sum of the geometric areas $Q_{n}^{(i)}$ in Fig. \ref{fig_aoiHarq}:
\begin{IEEEeqnarray}{rCl}
	\label{aoi_tr}
	\mathbb{E}[{h}_{n}(t)] = \lim_{T \to \infty} \frac{D}{T} \frac{1}{D} \sum_{i=1}^{I} Q_{n}^{(i)} &=&  \frac{\mathbb{E}[Q_{k,n}]}{\eta }, 
\end{IEEEeqnarray}
where $D$ denotes the number of successful deliveries until time $T$, and 
\begin{equation}
\eta \triangleq \mathbb{E}\left[S_{n,\mathsf{RR\_P}}^{(j)}\right]=\sum_{m=1}^N \mathbb{E}\left[K_{m}^{(i)}\right].
\end{equation} 
When $T$ goes to infinity, $D$ also goes to infinity. The last equality is based on the elementary renewal theorem \cite{cox67}. It then follows that
\begin{IEEEeqnarray}{rCl}
	\label{cons_cond}
	\mathbb{E}[{h}_{n}(t)] &=& \frac{1}{\eta} \mathbb{E}\left[S_{n,\mathsf{RR\_P}}^{(i)} K_n^{(i-1)} + \left(S_{n,\mathsf{RR\_P}}^{(i)}-1\right)\frac{S_{n,\mathsf{RR\_P}}^{(i)}}{2}\right] \nonumber\\
	&\overset{(a)}{=}& \frac{1}{\eta} \left(\mathbb{E}\left[S_{n,\mathsf{RR\_P}}^{(i)}\right]\mathbb{E}\left[ K_n^{(i-1)}\right] \right.\nonumber\\
	&& \left.+ \frac{1}{2} \left(\mathbb{E}\left[\left(S_{n,\mathsf{RR\_P}}^{(i)}\right)^2\right]-\mathbb{E}\left[S_{n,\mathsf{RR\_P}}^{(i)}\right]\right)\right) \nonumber\\
	& = & \mathbb{E}\left[ K_n^{(i-1)}\right] + \frac{1}{2\eta} \mathbb{E}\left[\left(\sum_{m=1}^NK_m^{(i)}\right)^2\right]-\frac{1}{2} \nonumber\\
	&\overset{(b)}{\le}& \mathbb{E}\left[ K_n\right] + \frac{1}{2\eta} \left(\sum_{m=1}^N \mathbb{E} \left[K_m^2\right] + \frac{N-1}{N}\eta^2\right)-\frac{1}{2}\nonumber\\
	&=& \mathbb{E}\left[ K_n\right] -\frac{\eta}{2N} + \frac{1}{2\eta} \sum_{m=1}^N \mathbb{E} \left[K_m^2\right] + \frac{\eta-1}{2}
\end{IEEEeqnarray}
where the equality $(a)$ is based on the fact that the number of transmission attempts during the $(i-1)$-th round is independent with the ones in the $i$-th round, and the inequality $(b)$ follows from the fact that for independent random variables $x_1,...,x_N$,
\begin{iarray}
	&& \mathbb{E}\left[\left(\sum_{i=1}^N x_i\right)^2\right] \nonumber\\
	&=& \sum_{i=1}^N \mathbb{E}\left[ x_i^2\right] + \sum_{i < j}^N 2 \mathbb{E}\left[ x_i\right]  \mathbb{E}\left[x_j\right] \nonumber\\
	&=& \sum_{i=1}^N \mathbb{E}\left[ x_i^2\right] + \left(\frac{1}{N} + 1-\frac{1}{N}\right)\sum_{i < j}^N 2 \mathbb{E}\left[ x_i\right]  \mathbb{E}\left[x_j\right] \nonumber\\
	&=& \sum_{i=1}^N \mathbb{E}\left[ x_i^2\right] + \frac{N-1}{N} \sum_{i=1}^N \mathbb{E}\left[ x_i\right]^2 +   \frac{N-1}{N}\sum_{i < j}^N 2 \mathbb{E}\left[ x_i\right]  \mathbb{E}\left[x_j\right] \nonumber\\
	&=& \sum_{i=1}^N \mathbb{E}\left[ x_i^2\right] + \frac{N-1}{N} \left(\sum_{i=1}^N \mathbb{E}\left[ x_i\right]\right)^2.
\end{iarray}
Note that after inequality $(b)$, we ignore the index of round for brevity. Now averaging over all terminals, we obtain
\begin{IEEEeqnarray}{rCl}
	\bar{\Delta}_{\mathsf{RR\_ P}} &=& \frac{1}{N} \sum_{n=1}^N \mathbb{E}[{h}_{n}(t)] \nonumber\\
	& \overset{(a)}{=} & \frac{1}{N} \sum_{n=1}^N \mathbb{E}\left[ K_n^{(i-1)}\right] + \frac{1}{2\eta} \mathbb{E}\left[\left(\sum_{m=1}^NK_m^{(i)}\right)^2\right]-\frac{1}{2} \nonumber\\
	&\le& \frac{\eta}{2N} + \frac{1}{2\eta} \sum_{m=1}^N \mathbb{E} \left[K_m^2\right] + \frac{\eta-1}{2},
\end{IEEEeqnarray}
wherein the equality $(a)$ gives \eqref{rrpExact} directly, and \eqref{rrpBound} readily follows given $\eta = N \bar{\mathbb{M}} \left[ g(\boldsymbol{\omega}_{i,n})\right]$. For the asymptotic results of \eqref{rrpAsym}, considering the HARQ models in \eqref{gr} and Lemma \ref{lmInterval}, we obtain respectively for two models,
\begin{IEEEeqnarray}{rCl}
	\bar{\Delta}_{\mathsf{RR\_ P},1} &\le& \frac{N+1}{2} \bar{\mathbb{M}} \left[ g(\boldsymbol{\omega}_{1,n})\right] + \frac{1}{2\eta} \sum_{m=1}^N \mathbb{E} \left[K_{1,m}^2\right] - \frac{1}{2} \nonumber\\
	&\le& \frac{N+1}{2} \bar{\mathbb{M}} \left[ g(\boldsymbol{\omega}_{1,n})\right] +  \frac{\sum_{m=1}^Np_{m,0}e^{p_{m,0}}}{\sum_{m=1}^N e^{p_{m,0}}} \nonumber\\
	&\le& \frac{N+1}{2} \bar{\mathbb{M}} \left[ g(\boldsymbol{\omega}_{1,n})\right] + 1. \nonumber
\end{IEEEeqnarray}
Given that 
\begin{equation}
	\bar{\mathbb{M}} \left[ g(\boldsymbol{\omega}_{1,n})\right] = \frac{1}{N} \sum_{m=1}^N \mathbb{E} \left[K_{1,m}\right] \le e,
\end{equation}
and that the left inequality of \eqref{rrpAsym} is straightforward, we can conclude the asymptotic results immediately. Similarly, 
\begin{IEEEeqnarray}{rCl}
	&& \bar{\Delta}_{\mathsf{RR\_ P},2} \nonumber\\
	&\le& \frac{N+1}{2} \bar{\mathbb{M}} \left[ g(\boldsymbol{\omega}_{2,n})\right] + \frac{1}{2\eta} \sum_{m=1}^N \mathbb{E} \left[K_{2,m}^2\right] - \frac{1}{2} \nonumber\\
	&\le& \frac{N+1}{2} \bar{\mathbb{M}} \left[ g(\boldsymbol{\omega}_{2,n})\right] + \frac{\sum_{m=1}^N \left[-\frac{\beta_m}{\alpha}\mathbb{E}[K_{2,m}]+\frac{\beta_m+1}{\alpha}-1\right]}{2\sum_{m=1}^N \mathbb{E} \left[K_{2,m}\right]} \nonumber\\
	&\overset{(a)}{\le}& \frac{N+1}{2} \bar{\mathbb{M}} \left[ g(\boldsymbol{\omega}_{2,n})\right] \nonumber\\
	&& + \frac{\sum_{m=1}^N \left[-\frac{\beta_m}{\alpha}  \left(1+\sqrt{\frac{\pi}{\alpha}}\right)p_{m,0} +\frac{1}{\alpha}-1\right]}{2\sum_{m=1}^N \mathbb{E} \left[K_{2,m}\right]} \nonumber\\
	&\overset{(b)}{\le}& \frac{N+1}{2} \bar{\mathbb{M}} \left[ g(\boldsymbol{\omega}_{2,n})\right] + \frac{1}{2\alpha},
\end{IEEEeqnarray}
wherein $\beta_m=-\log p_{m,0}$, $\alpha = -\frac{1}{2}\log\lambda$, the inequality $(a)$ follows from Lemma \ref{lmInterval} and $(b)$ is obtained by noting $\min_{m}\left[\mathbb{E} \left[K_{2,m}\right]\right] \ge 1$. With 
\begin{equation}
\bar{\mathbb{M}} \left[ g(\boldsymbol{\omega}_{2,n})\right] = \frac{1}{N} \sum_{m=1}^N \mathbb{E} \left[K_{2,m}\right] \le 2+\sqrt{\frac{\pi}{\alpha}},
\end{equation}
which is irrelevant with $N$, the conclusion follows immediately.

\section*{Acknowledgment}
This work was supported by the National Key R$\&$D Program of China (No. 2017YFE0121400), the program for Professor of Special Appointment (Eastern Scholar) at Shanghai Institutions of Higher Learning, and Shanghai Institute for Advanced Communication and Data Science (SICS).

\bibliographystyle{ieeetr}
\bibliography{aoi}
\end{document}